\documentclass[oneside,twocolumn]{article}

\usepackage{geometry}
\usepackage{fullpage}

\usepackage[utf8]{inputenc}
\usepackage{graphicx}
\usepackage{amsmath,mathtools}
\usepackage{amssymb}
\usepackage{amscd}
\usepackage{bm}
\usepackage{url}
\usepackage[super,sort&compress,comma]{natbib}
\usepackage{upgreek}
\usepackage{hyperref}
\usepackage{pdflscape}
\usepackage[font=small,labelfont=bf,labelsep=period]{caption}
\usepackage[version=3]{mhchem}
\usepackage{booktabs}
\usepackage{multirow}
\usepackage{threeparttable}
\usepackage{sectsty}
\usepackage{setspace}
\usepackage{authblk}
\usepackage{fancyhdr}
\usepackage{enumitem}

\usepackage[T1]{fontenc}
\usepackage{textcomp}
\usepackage[euler]{textgreek}

\makeatletter
  \def\@seccntformat#1{\csname the#1\endcsname.\ \ }
\makeatother

\sectionfont{\large}
\subsectionfont{\normalsize}
\subsubsectionfont{\normalsize}

\makeatletter
 \def\@biblabel#1{#1.}
\makeatother
\DeclarePairedDelimiter{\nbr}{(}{)}

\DeclarePairedDelimiter{\sbr}{[}{]}
\DeclarePairedDelimiter{\abs}{\lvert}{\rvert}
\DeclarePairedDelimiter{\ensav}{\langle}{\rangle}

\renewcommand*{\vec}[1]{\ensuremath{\bm{#1}}}

\newcommand*{\dd}{\mathrm{d}}
\newcommand*{\rsub}[1]{\ensuremath{_{\mathrm{#1}}}}
\newcommand*{\rsup}[1]{\ensuremath{^{\mathrm{#1}}}}

\newcommand*{\nga}{\ensuremath{N_{\mathrm{GA}}}}
\newcommand*{\pka}{\ensuremath{\mathrm{p}K_{\mathrm{a}}}}
\newcommand*{\degc}{\ensuremath{^{\circ}\mathrm{C}}}

\newcommand*{\inm}{\ensuremath{\mathrm{nm}^{-1}}}

\newcommand*{\us}{{\textmugreek}s}
\newcommand*{\ul}{{\textmugreek}l}

\newcommand*{\mM}{m\textsc{m}}
\newcommand*{\mgml}{\ensuremath{\mathrm{mg\,ml}^{-1}}}
\newcommand*{\perh}{\ensuremath{\mathrm{h}^{-1}}}

\newcommand*{\Hone}{\ce{^1H}}
\newcommand*{\Htwo}{\ce{^2H}}
\newcommand*{\Osvt}{\ce{^{17}O}}
\newcommand*{\rarrow}{\ensuremath{\rightarrow}}

\makeatletter
\def\@maketitle{%
  \newpage
  \begingroup
  \let \footnote \thanks
  \hrule height \z@        
    {\LARGE \bfseries \@title \par}%
    \vskip 6mm
    {\large
      \@author
    }%
  \par\endgroup
  \vskip 5mm 
  \vspace*{0cm}
}
\makeatother

\title{Structure and kinetics of chemically cross-linked protein gels from 
       small-angle X-ray scattering}

\author[1]{Shuji Kaieda}
\author[2]{Tom\'as S.\ Plivelic}
\author[1]{Bertil Halle}

\affil[1]{Department of Biophysical Chemistry, Lund University, 
          Lund, Sweden}
\affil[2]{MAX IV Laboratory, Lund University, 
          Lund, Sweden}

\date{}

\begin{document}

%
%
\twocolumn[
\begin{@twocolumnfalse}

\maketitle\thispagestyle{fancy}

%
%
\noindent Glutaraldehyde (GA) reacts with amino groups in proteins, forming 
intermolecular cross-links that, at sufficiently high protein concentration, can 
transform a protein solution into a gel.   
Although GA has been used as a cross-linking reagent for decades, neither the 
cross-linking chemistry nor the microstructure of the resulting protein gel have 
been clearly established.
Here we use small-angle X-ray scattering (SAXS) to characterise the 
microstructure and structural kinetics of gels formed by cross-linking of 
pancreatic trypsin inhibitor, myoglobin or intestinal fatty acid-binding 
protein. 
By comparing the scattering from gels and dilute solutions, we extract the 
structure factor and the pair correlation function of the gels.
The protein gels are spatially heterogeneous, with dense clusters linked by 
sparse networks. 
Within the clusters, adjacent protein molecules are almost in contact, but the 
protein concentration in the cluster is much lower than in a crystal. 
At the $\sim 1$~nm SAXS resolution, the native protein structure is unaffected 
by cross-linking. 
The cluster radius is in the range 10 -- 50~nm, with the cluster size determined 
mainly by the availability of lysine amino groups on the protein surface. 
The development of structure in the gel, on time scales from minutes to hours, 
appears to obey first-order kinetics. 
Cross-linking is slower at acidic pH, where the population of amino groups in 
the reactive deprotonated form is low. 
These results support the use of cross-linked protein gels in NMR studies of 
protein dynamics and for modeling NMR relaxation in biological tissue.   
\vspace{0.5cm}
\end{@twocolumnfalse}
]

%
%
\section{\label{sec:intro}Introduction}

Glutaraldehyde (GA; 1,5-pentanedial) has been widely used during the past 50 
years to immobilise and stabilise proteins through covalent intermolecular 
cross-links.\cite{Walt1994,Migneault2004}
This bifunctional reagent has been used as a fixative in studies of cell or 
tissue ultrastructure,\cite{Sabatini1963,Hopwood1972} to stabilise protein 
crystals for X-ray diffraction,\cite{Quiocho1964,Wine2007} and to characterise 
the quaternary structure of proteins in solution.\cite{Hall2005,Fadouloglou2008}
Applications of GA cross-linking in immunochemistry and biotechnology include 
protein immobilisation on solid carriers, \textit{e.g.}, in affinity 
chromatography and biosensors, as well as carrier-free immobilisation of 
enzymes in solution, in amorphous precipitates, or in crystals for use as 
industrial biocatalysts.\cite{Habeeb1967,Payne1973,Roy2004,Sheldon2007,
Talekar2013}   

The present study was motivated by yet another application of protein 
cross-linking: water NMR studies of biological systems. 
In a protein solution, all anisotropic nuclear spin couplings are averaged out 
by protein tumbling.
As a result, the longitudinal relaxation of the water (\Hone, \Htwo\ or \Osvt)  
magnetisation only reports on molecular motions faster than the protein's 
tumbling time (typically, $\sim 10$~ns). 
Protein cross-linking profoundly alters the NMR conditions, allowing motions on 
time scales up to hundreds of \us\ to influence the relaxation.
In the NMR context, cross-linked protein gels were first used as model systems 
for biological tissue,\cite{Grad1990,Bryant1991,Zhou1994,Koenig1993,Sunde2009} 
wherein the proteins are largely immobilised,\cite{Persson2008b} in efforts to 
elucidate the molecular determinants of the water \Hone\ relaxation that governs 
contrast in magnetic resonance images of soft tissue.   
More recently, cross-linked protein gels have been used in \Htwo\ and \Osvt\ 
magnetic relaxation dispersion (MRD) studies of intermittent protein dynamics on 
the ns -- \us\ time scale.\cite{Persson2008a,Kaieda2013a}      

GA reacts primarily with the protein's amino groups, in lysine side-chains and 
at the N-terminus, although minor involvement of other residues (arginine, 
histidine, tyrosine and cysteine) has been reported.\cite{Richards1968,
Habeeb1968,Monsan1975,Okuda1991,Salem2010}
Despite extensive study, the details of the reaction mechanism remain poorly 
understood.\cite{Walt1994,Migneault2004}
Besides the dialdehyde, an aqueous solution of GA contains several species in 
equilibrium, including hemihydrate, dihydrate, cyclic hemiacetal, polymeric 
forms of the hemiacetal and various aldol condensation adducts.\cite{Hardy1969,
Korn1972,Whipple1974,Margel1980,Tashima1991,Kawahara1992}
These equilibria, which depend on pH, temperature and concentration, may account 
for the efficiency of GA as a cross-linking agent by allowing it to form linkers 
of variable length. 

In quantitative MRD studies of protein dynamics, the cross-links should ideally 
inhibit protein tumbling without affecting the internal (conformational) 
dynamics of the protein. 
A necessary condition for this is that the protein structure is unaffected by GA 
cross-linking, except locally at the chemically modified residues. 
For protein crystals, X-ray diffraction demonstrates that the structural 
perturbation caused by cross-linking is indeed local.\cite{Wine2007,Salem2010} 
For cross-linked protein gels, the evidence is less direct, but the limited 
results available so far have not revealed any significant differences in 
internal protein dynamics between gel and solution.\cite{Persson2008a} 

The protein gels used in MRD studies are formed by adding GA to protein 
solutions at concentrations where the protein molecules are separated by several 
water layers.\cite{Persson2008a,Kaieda2013a} 
But even if the protein is amply hydrated on average, the protein molecules may 
not be uniformly distributed. 
Cross-linking may well produce a gel structure with dense tightly cross-linked 
protein clusters connected by more dilute weakly cross-linked networks. 
Even if such spatial heterogeneity has little effect on the internal dynamics, 
water dynamics in the first hydration layer on the protein surface would be
affected.\cite{Persson2008a,Kaieda2013a}          

To our knowledge, the structure of chemically cross-linked protein gels has not 
been examined directly. 
Such studies would have implications for the interpretation of MRD data from 
cross-linked protein gels and, more generally, would further our understanding 
of protein cross-linking by GA. 
A technique suitable for this task is small-angle X-ray scattering 
(SAXS),\cite{Guinier1955,Kratky1982} which can 
provide information about gel structure via the structure factor, essentially 
the Fourier transform of the protein--protein pair correlation function, as 
well as about the integrity of the protein's tertiary structure via the form 
factor.

In protein science, SAXS has proven useful for determining the low-resolution 
structure of monomeric and oligomeric proteins in dilute 
solution,\cite{Koch2003,Jacques2010} but this technique has also been used to 
study protein--protein interactions in more concentrated 
solutions\cite{Haussler2002,Zhang2007,Cardinaux2011} and to obtain structural 
information about more complex protein systems, such as casein 
micelles,\cite{Kruif2012} gluten films\cite{Kuktaite2011} and gels of heat 
denatured proteins.\cite{Pouzot2004}

Here, we report SAXS data for GA cross-linked gels of three proteins:
bovine pancreatic trypsin inhibitor (BPTI), equine skeletal muscle myoglobin 
(Mb) and rat intestinal fatty acid-binding protein (IFABP). 
MRD studies of \us\ protein dynamics and internal water exchange in these 
protein gels have already been performed (BPTI and Mb)\cite{Persson2008a,
Kaieda2013a} or are currently underway (IFABP).\cite{Kaieda2013e} 
For each protein, we analyse the scattering intensity profiles in terms of the 
inhomogeneous protein distribution in the gel.  
We also report time-resolved SAXS measurements of the cross-linking kinetics.

%
%
\section{\label{sec:exp}Experimental}

%
%
\subsection{\label{subsec:sample}Sample preparation}

%
%
\subsubsection{\label{subsubsec:bpti_prep}BPTI.~~}

{\small
BPTI (trade name Trasylol, batch 9104; 97~\% purity by HPLC) was obtained from 
Bayer HealthCare AG (Wuppertal, Germany). 
To remove residual salt, the protein was extensively dialysed against MilliQ 
water (Millipore) and then lyophilised.
}

%
%
\subsubsection{\label{subsubsec:mb_prep}Mb.~~}

{\small
Equine skeletal muscle Mb ($\geq 95~\%$) was purchased from Sigma. 
The protein was further purified by cation-exchange chromatography (SP sepharose; 
GE Healthcare), dialysed against MilliQ water and lyophilised.
}

%
%
\subsubsection{\label{subsubsec:ifabp_prep}IFABP.~~}

{\small
The gene encoding rat IFABP was codon optimised for expression in 
\textit{Escherichia coli} and synthesised by DNA2.0 (Menlo Park, CA, USA). 
The synthetic DNA was inserted into the pNIC28-Bsa4 plasmid\cite{Savitsky2010} 
for expression. 
The expression construct yields a fusion protein containing, from the N-terminus, 
the His$_6$-tag, tobacco etch virus (TEV) protease cleavage site and IFABP. 
The fusion protein was over-expressed using \textit{E.\ coli} TUNER(DE3) strain 
(Novagen) in Terrific Broth (Difco). 
After harvesting, the bacterial cells were suspended in a lysis buffer (50~\mM\ 
sodium phosphate, 300~\mM\ NaCl, 10~\mM\ imidazole, pH~8.0) and homogenised 
by French press. 
The cell lysate was ultracentrifuged and the supernatant was subjected to 
His$_6$-tag affinity chromatography (HisTrap; GE Healthcare). 
The His$_6$-tag of the fusion protein was then cleaved off by TEV protease. 
After the protease digestion, the His$_6$-tag and the protease were removed 
by passing the solution through the HisTrap column and the flow-through fraction 
containing IFABP was collected. 
IFABP was then delipidated by using a Lipidex-1000 (Perkin Elmer) column. 
The IFABP solution was then dialysed against MilliQ water and lyophilised.
}

%
%
\subsubsection{\label{subsubsec:saxs_sample}SAXS samples.~~}

{\small
The lyophilised proteins were dissolved in MilliQ water (cross-linked BPTI and 
solution samples for all proteins) or in a buffer solution (50~\mM\ PIPES for 
cross-linked Mb, 50~\mM\ sodium phosphate for cross-linked IFABP). 
The solution was then centrifuged at 13\,000~rpm for 3~min to remove any 
insoluble proteins. 
To prepare cross-linked samples, the protein solution was supplemented with 
25~\% glutaraldehyde solution (Sigma). 
After vigorous mixing, the solution was transferred to a 1.5~mm o.d.\ 
borosilicate capillary (Hilgenberg GmbH) where the cross-linking reaction 
proceeded at 6~$\degc$. 
Approximately 50~\ul\ of the solution was reserved for pH measurement. 
SAXS measurements were also performed on protein solutions without GA.
The pH of these solution samples was adjusted to match that of the corresponding 
cross-linked sample by adding either HCl or NaOH.
}

%
%
\subsection{\label{subsec:saxsexp}SAXS measurements}

{\small
SAXS experiments were carried out at the I911-4 beamline\cite{Labrador2013} of 
the MAX-lab synchrotron using a wavelength of 0.91~\AA. 
The sample, contained either in a capillary (gels) or in a flow-through cell 
(solutions), was maintained at 20~$\degc$ or, for the kinetics experiments, at 
6~$\degc$. 
For each protein sample, a pure solvent sample (MilliQ water or buffer solution) 
was also measured.
Two-dimensional SAXS images were recorded with a PILATUS 1M detector (Dectris) 
with an exposure time of 10~s (kinetics series for Mb and IFABP) or 60~s (in all 
other cases). 
Control measurements were performed to ensure that the results were not 
compromised by radiation damage.
The scattering vector $q$ range ($q = 4\pi\sin\theta/\lambda$, where $\lambda$ 
is the wavelength and $2\theta$ is the scattering angle) was calibrated with a 
silver behenate sample. 
Reported scattering profiles $I(q)$ were obtained as the difference of the 
azimuthally averaged SAXS 2D images from sample and solvent.
}

%
%
\subsection{\label{subsec:saxsanal}SAXS analysis}

{\small
For a sample of $N\rsub{P}$ identical protein molecules of volume $V\rsub{P}$ 
contained in a volume $V$, the scattering intensity in the decoupling 
approximation, where the orientation of a protein molecule is taken to be 
independent of its position and the configuration of other protein molecules, 
can be factorised as\cite{Guinier1955,Kratky1982,Pedersen1997}
\begin{equation}
  I\nbr*{q} = n\rsub{P} \, \nbr*{V\rsub{P} \, \Delta\rho}^2 \, P\nbr*{q} \, 
  S\nbr*{q}\ ,
  \label{eq:iq}
\end{equation}
where $n\rsub{P} = N\rsub{P}/V$ is the protein number density and the scattering 
contrast $\Delta\rho$ is the protein--solvent electron density difference. 
The scattering from an isolated protein molecule is described by the form 
factor\cite{Guinier1955,Kratky1982,Pedersen1997} 
\begin{equation}
  P\nbr*{q} = \ensav*{\abs*{\frac{1}{V\rsub{P}} \int_{V\rsub{P}} \dd\vec{r} \, 
  \exp\nbr*{-i \, \vec{q} \cdot \vec{r}}}^2}\ ,
  \label{eq:pq}
\end{equation}
while information about the gel structure is contained in the structure factor
\begin{equation}
  S(q) = \sum_{k=1}^{N\rsub{P}} \, 
  \ensav*{\exp\sbr*{-i \, \vec{q} \cdot \nbr*{\vec{r}_1 - \vec{r}_k}}}\ ,
  \label{eq:sq}
\end{equation}
where $\ensav*{\ldots}$ denotes equilibrium configurational averaging for the 
isotropic system.
Strictly speaking, the structure factor in Eq.~\eqref{eq:iq} should be regarded 
as an effective structure factor $\bar{S}\nbr*{q}$.\cite{Pedersen1997} 
For the samples studied here, the difference between $\bar{S}\nbr*{q}$ and 
$S(q)$ due to non-spherical protein shape is likely to be small.

According to Eq.~\eqref{eq:iq}, the structure factor $S\nbr{q;\,n\rsub{P}}$ for 
a protein gel at concentration $n\rsub{P}$ can be obtained from the 
corresponding SAXS intensity $I\nbr*{q;\,n\rsub{P}}$ and the intensity 
$I\nbr*{q;\,n\rsub{P}^0}$ measured from a solution of the same protein at a 
concentration $n\rsub{P}^0$ sufficiently low that $S\nbr{q;\,n\rsub{P}^0} = 1$: 
\begin{equation}
  S\nbr{q;\,n\rsub{P}} = \frac{I\nbr*{q;\,n\rsub{P}}}{n\rsub{P}} 
  \times \frac{n\rsub{P}^0}{I\nbr*{q;\,n\rsub{P}^0}}\ .
  \label{eq:sq_exp}
\end{equation}
This approach neglects, in the $q$ range considered, the direct contribution to 
$P\nbr*{q}$ from GA as well as any effects of the cross-links on the protein 
structure via $P\nbr*{q}$ and $\Delta\rho$. 

For each protein, the quantity $I\rsub{P}\nbr*{q;\,n\rsub{P}^0}$ in 
Eq.~\eqref{eq:sq_exp}, hereafter referred to as the apparent form factor (AFF), 
was obtained by merging solution SAXS profiles recorded at two different protein 
concentrations, as shown in Fig.~\ref{fig:formfactor}.
The two profiles were first superimposed in an intermediate $q$ window ($q = 1.9 
\pm 0.3$, $1.5 \pm 0.1$ or $1.7 \pm 0.1~\inm$ for BPTI, Mb or IFABP, 
respectively), indicated by vertical dashed lines in Fig.~\ref{fig:formfactor}, 
and a hybrid profile was constructed by merging the low-$q$ part of the lower 
concentration profile with the high-$q$ part of the higher concentration profile. 
At low $q$, where the SAXS profile is sensitive to protein--protein correlations, 
we thus use data from the most dilute solution, while, at high $q$, we benefit 
from the better signal-to-noise ratio of the data from the more concentrated 
solution.    
The merged profile was then smoothed with the aid of a Savitzky--Golay 
filter.\cite{Savitzky1964} 
For BPTI and Mb, a linear regression on a Guinier plot\cite{Guinier1955} was 
performed in a low-$q$ window, indicated by vertical dotted lines in 
Fig.~\ref{fig:formfactor}, and used to extrapolate the AFF to $q = 0$.
Finally, the AFF was obtained by scaling the merged profile by a constant factor 
that minimises the difference between the merged solution profile and the gel 
profile in a high-$q$ range (2.95 -- 3.05, 2.9 -- 3.1 or 3.9 -- 4.1~$\inm$ for 
BPTI, Mb or IFABP, respectively), where we expect that $S(q) = 1$. 
This scaling, which ensures that the gel structure factor derived from 
Eq.~\eqref{eq:sq_exp} tends to 1 at the highest $q$, is justified by the 
different sample containers used for gels (capillary) and solutions 
(flow-through cell).

For an isotropic sample of identical protein molecules, the structure factor 
is related to the Fourier transform of the protein--protein pair correlation 
function (PCF), $g\nbr*{r}$. After angular integration, one 
obtains\citep{Hansen1986}  
\begin{equation}
  S\nbr*{q} \:=\: 1 + \frac{4 \pi\,n\rsub{P}}{q} \int_0^\infty \dd r \, 
  r \sin\nbr*{qr} \, \sbr*{g\nbr*{r} - 1}\ .
  \label{eq:sq_ft}
\end{equation}
This relation may be inverted to obtain the PCF from the structure factor as
\begin{equation}
  g\nbr*{r} \:=\: 1 + \frac{1}{2 \pi^{\,2}\,n\rsub{P}\,r} \int_0^\infty \dd q 
  \, q \sin\nbr*{qr} \, \sbr*{S\nbr*{q} - 1}\ .
  \label{eq:gr_ft}
\end{equation}

To obtain the PCF from Eq.~\eqref{eq:gr_ft}, the structure factor, which has 
been experimentally determined in a finite $q$ range, must be extended to higher 
and lower $q$.
First, a value $q\rsub{max}$ is selected, above which $S\nbr*{q}$ is set equal 
to 1. 
This value, $q\rsub{max} = 3.0$, 3.0 or 4.0~$\inm$ for BPTI, Mb or IFABP, 
respectively, is taken as the midpoint of the $q$ range used to scale the AFF. 
Then a value $q\rsub{min}$ is selected (0.09, 0.065 or 0.075~$\inm$ for BPTI, Mb 
or IFABP, respectively), below which $S\nbr*{q}$ is extrapolated by fitting a 
quadratic polynomial to the 20 data points (covering $\sim 0.07~\inm$) just above 
$q\rsub{min}$. 
Using cubic spline interpolation, we resample $S(q)$ with fixed spacing 
$\Delta q$. 
Finally, we zero-fill $S(q) - 1$ to obtain a real-space resolution 
$\Delta r = 0.065$~nm.

It follows from Eq.~\eqref{eq:gr_ft} that even a small deviation of $S(q)$ from 
1 at high $q$ has a large effect on $g(r)$ at small $r$. Our protocol of 
setting $S(q) = 1$ above $q\rsub{max}$ thus causes $g(r)$ to be negative at 
small $r$ (Fig.~\ref{fig:gr_orig}). 
This unphysical feature is removed by forcing $g(r) = 0$ whenever the transform 
produces negative values.  

From the PCF, we compute the running coordination number as
\begin{equation}
  N\nbr*{r} \:=\: 4 \pi\,n\rsub{P} \int_0^r \dd r^\prime \, r^{\prime\,2} 
  g\nbr*{r^\prime}\ .
  \label{eq:nr}
\end{equation}
}

%
%
\section{\label{sec:results}Results and discussion}

%
%
\subsection{\label{subsec:structure}Structure factor and pair correlation 
function}

\begin{figure}[!t]
  \centering
  \includegraphics[viewport=0 0 231 349]{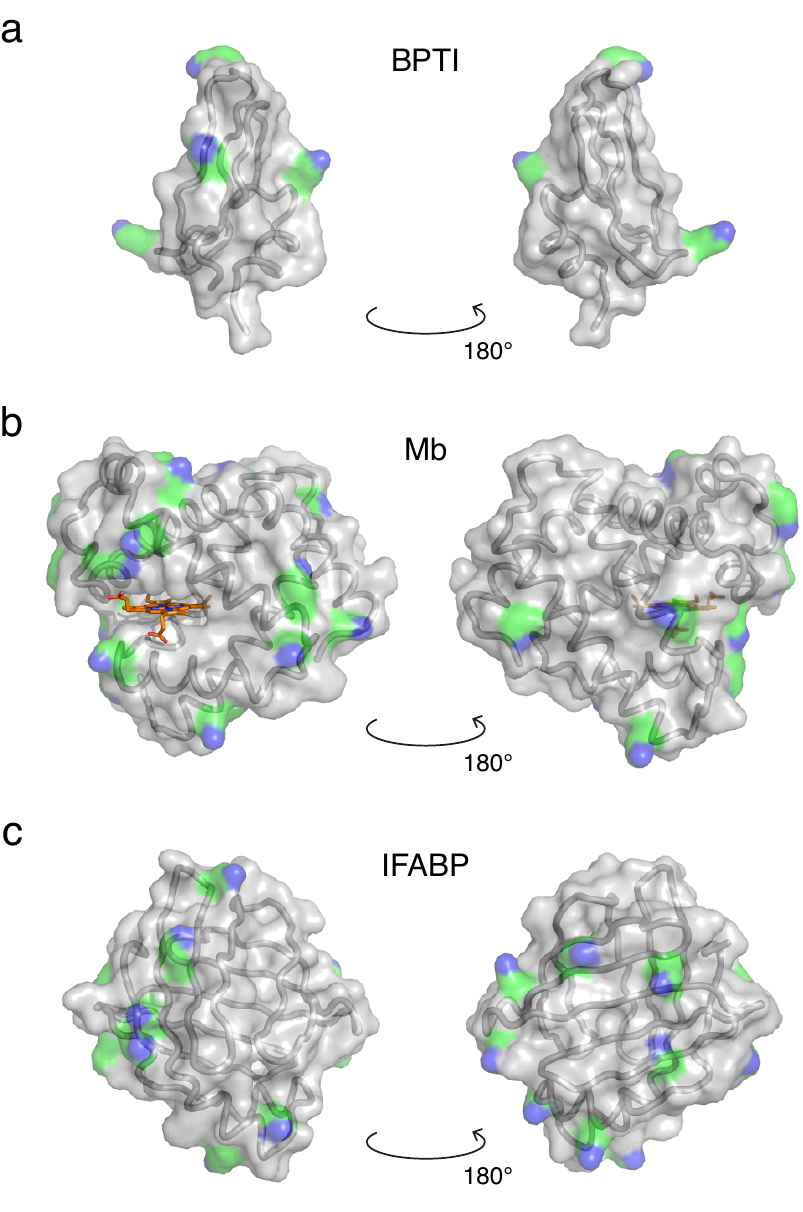}
  \caption{\label{fig:struct}Crystal structures of BPTI (PDB ID 1bpi\cite{Parkin1996}), 
           Mb (1wla\cite{Maurus1997}) and IFABP (1ifc\cite{Scapin1992}).
           Ribbon and surface representations are superimposed, while the haem 
           group of Mb is shown in a stick representation (C, N, O and Fe atoms 
           coloured orange, blue, red and brown, respectively).
           Lysine side-chains are coloured by element (C, green; N, blue).}
\end{figure}

Crystal structures\cite{Parkin1996,Maurus1997,Scapin1992} of the three 
investigated proteins, BPTI, Mb and IFABP, are shown in Fig.~\ref{fig:struct}. 
BPTI has a pear-like shape with principal dimensions of 2.1 and 3.4~nm and 
volume $V\rsub{P} =$ 7.792~nm$^3$. 
The shape of Mb is oblate-like, with principal dimensions 2.6 and 4.0~nm and 
$V\rsub{P} =$ 21.67~nm$^3$. 
IFABP is also oblate-like with principal dimensions 2.5 and 3.6~nm and 
$V\rsub{P} =$18.79~nm$^3$.
The effective diameter, $\sigma\rsub{P}$, for a sphere of volume $V\rsub{P}$ and 
the number, $N\rsub{Lys}$, of lysine residues per protein are listed in 
Table~\ref{tab:gel}, which also presents relevant gel sample characteristics 
such as the protein volume fraction $\phi\rsub{P}$ and the centre-to-centre 
separation $d\rsub{PP}$ between protein molecules assuming a cubic lattice.
The protein concentration, pH and GA/protein mole ratio, $N\rsub{GA}$, were 
chosen to match gel samples used in MRD studies.\cite{Persson2008a,Kaieda2013a,
Kaieda2013e}

\begin{table*}[!t]
  \centering
  \begin{threeparttable}
    \caption{\label{tab:gel}Characteristics of protein gel samples.}
    \small
    \begin{tabular}{lcccccccccc}
      \toprule
      \multirow{2}{*}{Protein} & \multirow{2}{*}{$N\rsub{Lys}$\tnote{a}} 
      & $\sigma\rsub{P}$\tnote{b} & $C\rsub{P}$ & $\phi\rsub{P}$\tnote{c} 
      & $d\rsub{PP}$\tnote{d} & \multirow{2}{*}{pH\,\tnote{e}} 
      & \multirow{2}{*}{$Z$\,\tnote{f}} & \multirow{2}{*}{$\nga$} 
      & $\xi$\,\tnote{g} & $k\rsub{CL}$\tnote{h} \\
      & & (nm) & (\mM\ / $\mgml$) & (\%) & (nm) & & & & (nm) & ($\perh$) \\
      \midrule
      BPTI  &  4 & 2.46 & 16.5 / 108  & 7.78 & 5.21 & 4.1 & $+7.3$ & 30.0 
            & 16 & $0.115 \pm 0.002$ \\[1ex]
      Mb    & 19 & 3.46 & 1.62 / 28.5 & 2.11 & 11.3 & 6.8 & $+3.7$ & 30.0 
            & 157 & $18.2 \pm 0.1$ \\[1ex]
      IFABP & 15 & 3.30 & 3.83 / 59.4 & 4.33 & 8.50 & 7.0 & $+0.2$ & 30.0 
            & 52 & $9.65 \pm 0.05$ \\
      \bottomrule
    \end{tabular}
    \footnotesize
    \begin{tablenotes}[flushleft,para]
      \item[a] Number of lysine residues.
      \item[b] Effective protein diameter.
      \item[c] Protein volume fraction.
      \item[d] Centre-to-centre separation in a cubic lattice, 
               $d\rsub{PP} = \sigma\rsub{P}
               \nbr*{\phi\rsub{cp}/\phi\rsub{P}}^{1/3}$ with 
               $\phi\rsub{cp} = \pi/(3\sqrt{2}) \approx 0.7405$.
      \item[e] pH measured in the cross-linked gel.
      \item[f] Net protein charge at given pH, calculated with standard $\pka$ 
               values and unmodified lysines. 
      \item[g] Correlation length (Sec.~\ref{subsec:description}).
      \item[h] Cross-linking rate at 6~$\degc$ at $q = 0.1~\inm$.
    \end{tablenotes}
  \end{threeparttable}
  \normalsize
\end{table*}

\begin{figure}[!t]
  \centering
  \includegraphics[viewport=0 0 209 428]{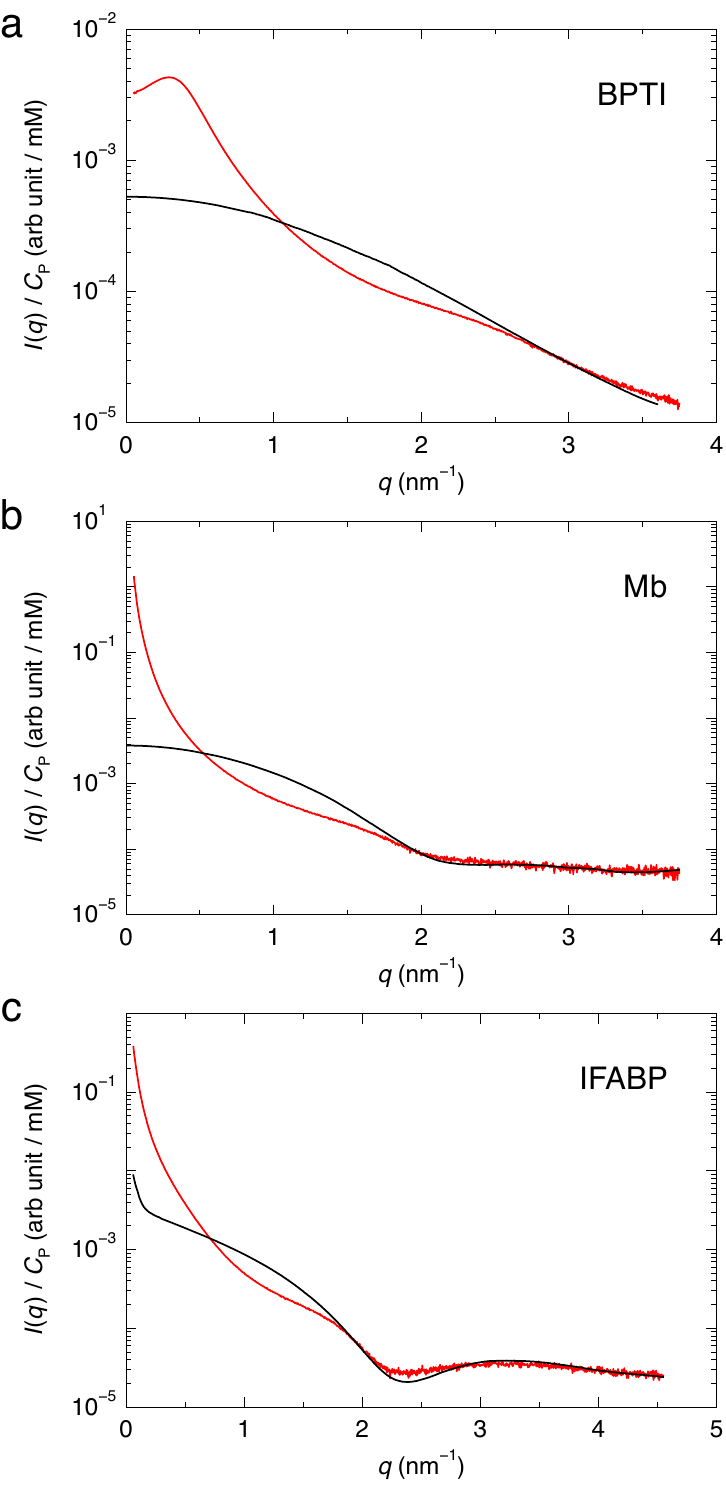}
  \caption{\label{fig:profile}Concentration normalised scattering profile from 
           cross-linked BPTI (\textbf{a}), Mb (\textbf{b}) and IFABP 
           (\textbf{c}). 
           The apparent form factor, derived from solution SAXS profiles as 
           described in Sec.~\ref{subsec:saxsanal}, is also shown for each 
           protein (black curve).}
\end{figure}

The concentration normalised (gel $-$ solvent) scattering profiles for the three 
protein gels are shown in Fig.~\ref{fig:profile}. 
The structure factor, $S\nbr*{q}$, was deduced from Eq.~\eqref{eq:sq_exp} using 
an apparent form factor (AFF), $I\nbr*{q;\,n\rsub{P}^0}$, constructed from SAXS 
profiles recorded on protein solutions at two concentrations 
(Sec.~\ref{subsec:saxsanal}, Figs.\ \ref{fig:profile} and~\ref{fig:formfactor}). 
For IFABP, the solution SAXS profile, and thus the AFF, increases sharply below 
$q \approx 0.2~\inm$ even at the lowest concentration (0.5~\mM), indicative of 
protein aggregation.  
A crude analysis shows that this feature in the scattering profile can be 
rationalised by a very small fraction ($\sim 10^{-5}$) of the IFABP molecules 
existing in large aggregates (effective diameter $\sim 10\,\sigma\rsub{P}$). 
For each protein, the AFF constructed from solution SAXS profiles at two 
concentrations agrees well (at $q \geq 1~\inm$ for IFABP) with a 
\textsc{crysol}\cite{Svergun1995} fit based on the crystal structure of the 
corresponding protein. 
We therefore conclude that the proteins are essentially monomeric form in our 
solution samples.

The close agreement between the SAXS profiles from the Mb and IFABP gels with 
the corresponding AFFs at high $q$ (Fig.~\ref{fig:profile}b,c), where we expect 
that $S(q) = 1$, indicates that the only effect of GA is to induce 
protein--protein correlations. 
The protein structure thus appears to be the same in solution and gel.
We note that although the AFF has been scaled to agree with the gel profile at 
high $q$ (Sec.~\ref{subsec:saxsanal}), this scaling does not alter the shape of 
the AFF.
For BPTI, the AFF cannot be scaled to superimpose with the gel profile over an 
extended high-$q$ range (Fig.~\ref{fig:profile}a), presumably because the 
$S(q) = 1$ limit is not reached in the investigated $q$ range for this small 
protein.
Another consequence of the limited $q$ range is that the dip in the AFF, which 
reflects the size and shape of the protein molecule, is observed for Mb and 
IFABP (at $q \approx 2.1$ and $2.3~\inm$, respectively; see 
Fig.~\ref{fig:profile}b,c), but not for the smaller BPTI. 
Using the program \textsc{crysol}\cite{Svergun1995} to compute the form factor 
from the BPTI crystal structure 1bpi,\cite{Parkin1996} we find a shallow dip at 
$q \approx 4.5~\inm$. 
Nevertheless, in the subsequent analysis of the BPTI profile, we postulate that 
$S\nbr*{q} = 1$ for $q \ge 3.0$~$\inm$.

\begin{figure}[!t]
  \centering
  \includegraphics[viewport=0 0 207 428]{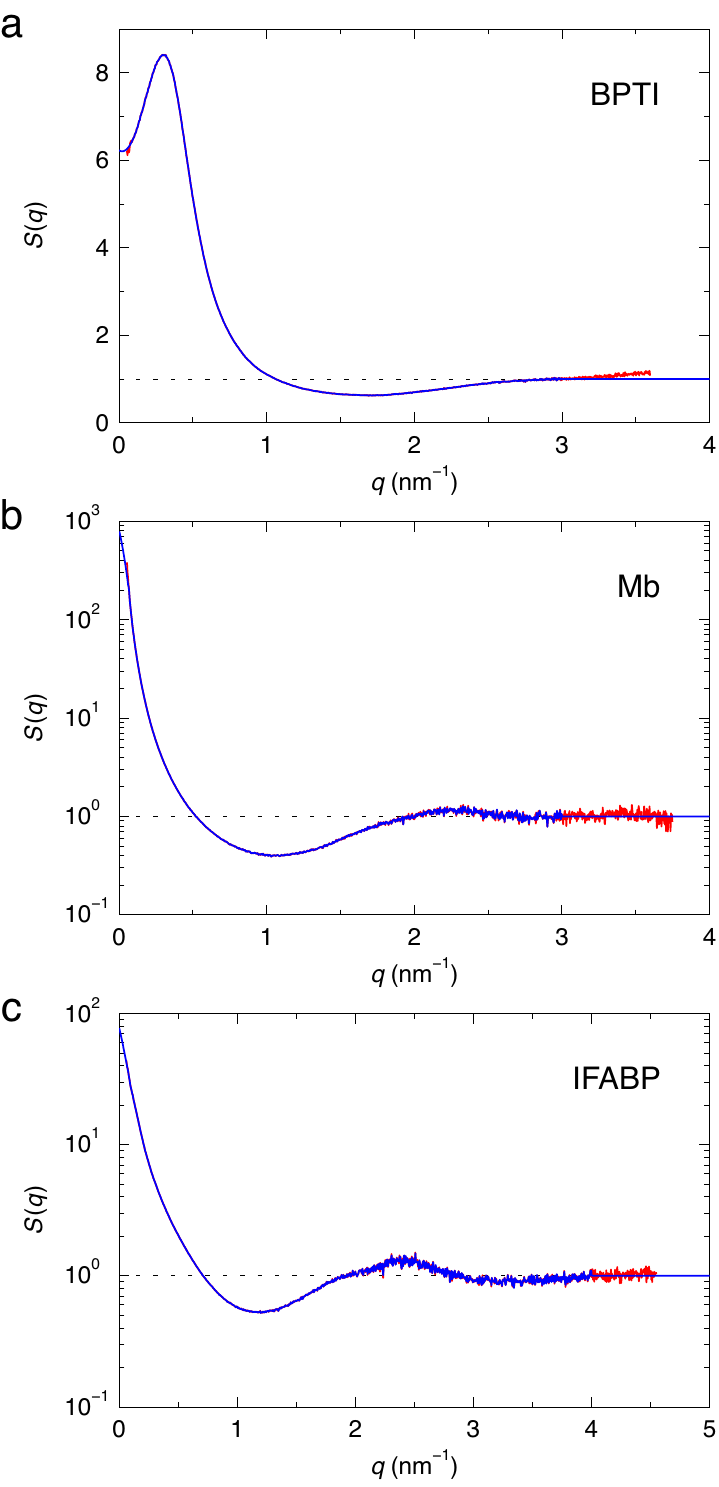}
  \caption{\label{fig:sq}Structure factor, $S\nbr*{q}$, for cross-linked BPTI 
           (\textbf{a}), Mb (\textbf{b}) and IFABP (\textbf{c}), obtained from 
           the profiles in Fig.~\ref{fig:profile} according to 
           Eq.~\eqref{eq:sq_exp} (red) and after truncation at high $q$ and 
           extension to high and low $q$ (blue). Note the linear scale in 
           (\textbf{a}).}
\end{figure}

The structure factors for the three protein gels, deduced in this way, are shown 
in Fig.~\ref{fig:sq}. 
For BPTI, $S(q)$ shows a pronounced maximum at $q \approx 0.3~\inm$, already 
evident in the gel profile (Fig.~\ref{fig:profile}a).
In contrast, for Mb and IFABP, $S(q)$ decreases monotonically with $q$ from the 
lowest examined $q$ value ($0.06~\inm$) up to $q \approx 1~\inm$ 
(Fig.~\ref{fig:sq}b,c).

\begin{figure}[!t]
  \centering
  \includegraphics[viewport=0 0 204 428]{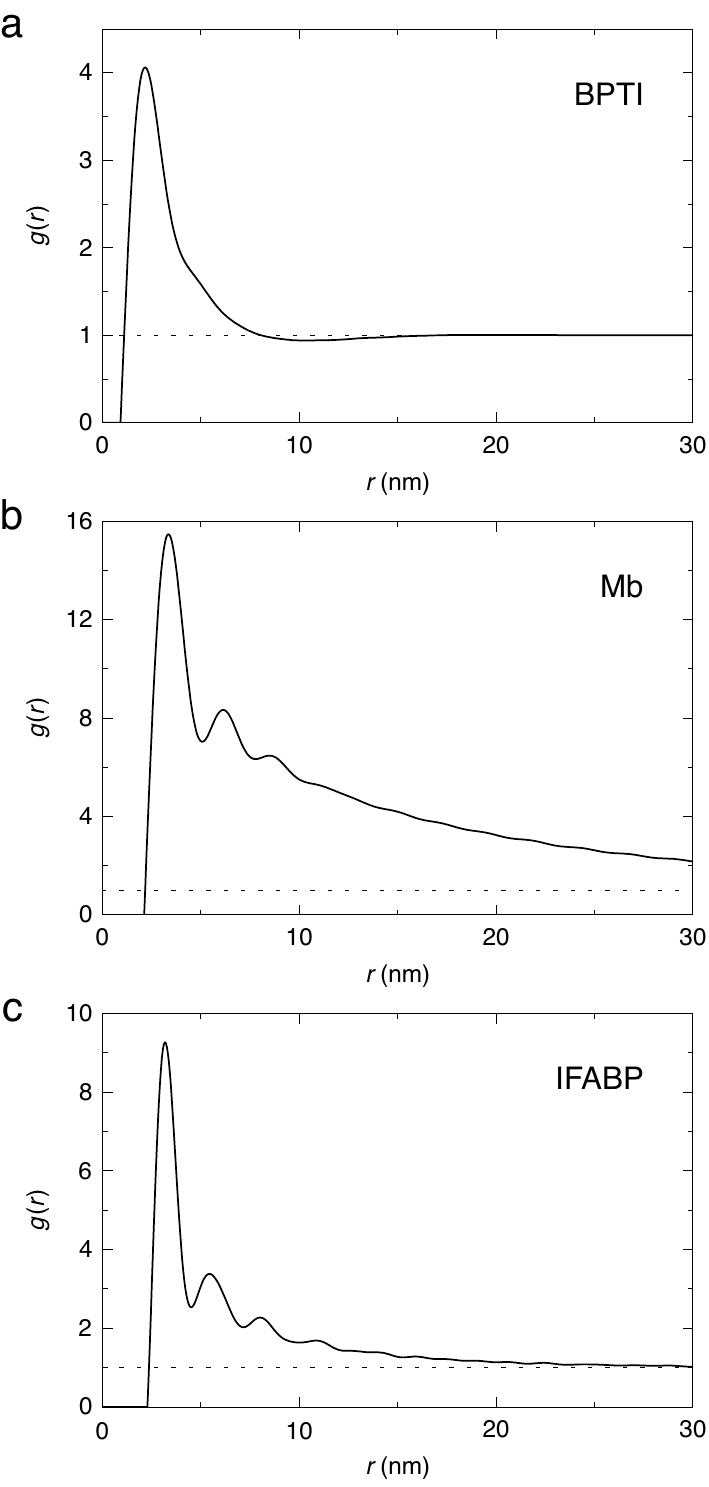}
  \caption{\label{fig:gr} Pair correlation function, $g\nbr*{r}$, for 
           cross-linked BPTI (\textbf{a}), Mb (\textbf{b}) and IFABP 
           (\textbf{c}), obtained by Fourier transformation of $S\nbr*{q}$ (blue 
           curves in Fig.~\ref{fig:sq}).}
\end{figure}

To characterise the gel microstructure in real space, we transform the structure 
factor into a pair correlation function (PCF), $g\nbr{r}$, with the aid of 
Eq.~\eqref{eq:gr_ft}, implemented as a fast sine transform. 
Before the transformation, $S(q)$ is truncated at high $q$ and extended to high 
and low $q$ as described in Sec.~\ref{subsec:saxsanal}. 
The modified $S(q)$ is included in Fig.~\ref{fig:sq} as a blue curve and the 
resulting PCF is shown in Fig.~\ref{fig:gr}.
Because we set $S(q) = 1$ at high $q$, the transformation yields negative $g(r)$ 
values at small $r$ (Fig.~\ref{fig:gr_orig}).
If we force $g(r) = 0$ at these small $r$ values and then inverse transform the 
corrected PCF according to Eq.~\eqref{eq:sq_ft}, we find that the 
back-calculated structure factor differs very little from the original one 
(Fig.~\ref{fig:sq_back}).
(However, a significant deviation is seen for Mb at $q \approx 1~\inm$.) 
This finding is consistent with the expectation that our SAXS data (with 
$q \le 4~\inm$) are insensitive to short-range ($r \lesssim 2$~nm) structural 
features. 
       
The PCF reflects the static spatial correlations in the sample, regardless of 
the origin of these correlations (cross-links or other interactions), and it can 
therefore be obtained from the structure factor without any assumptions about 
the structure of the gel (apart from isotropy). 
The analysis in Sec.~\ref{subsec:description} indicates that the protein gel is 
inhomogeneous on the length scale probed by our SAXS data, with dense protein 
clusters connected by less dense regions. 
This inhomogeneity complicates the interpretation of $g\nbr*{r}$. 
One possible approach would then be to model the protein gel as a mixture of 
clustered and non-clustered protein molecules. 
Such an approach would, however, introduce more parameters than can be justified 
by the data. 
Instead, we assume that the observed scattering is strongly dominated by 
clustered proteins so that the contribution from non-clustered proteins can be 
neglected.

\begin{figure}[!t]
  \centering
  \includegraphics[viewport=0 0 205 428]{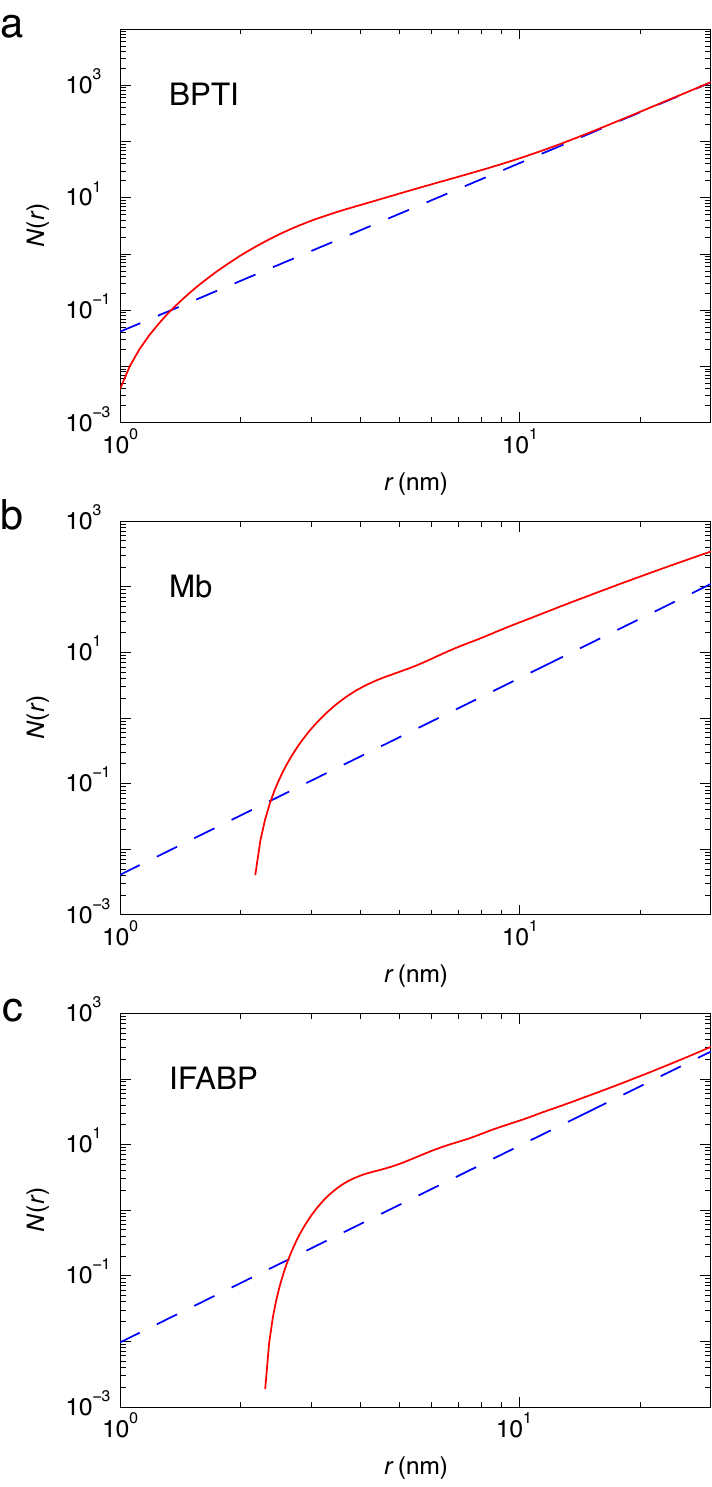}
  \caption{\label{fig:rcn}Running coordination number, $N\nbr*{r}$, for 
           cross-linked BPTI (\textbf{a}), Mb (\textbf{b}) and IFABP 
           (\textit{c}), obtained from the corrected PCF in Fig.~\ref{fig:gr} 
           (red solid curve) or by assuming a uniform ($g\nbr*{r} \equiv 1$) 
           protein distribution (blue dashed line).}
\end{figure}

The PCF is then used, along with Eq.~\eqref{eq:nr}, to obtain the running 
coordination number, $N\nbr*{r}$, in the protein gel (Fig.~\ref{fig:rcn}). 
For comparison, we show $N\nbr*{r}$ for a uniform protein distribution at the 
same protein concentration as in the corresponding gel.

%
%
\subsection{\label{subsec:description}Microstructure of protein gels}

%
%
\subsubsection{\label{subsubsec:bpti}BPTI.~~}

Among the three investigated protein gels, only the BPTI gel produces a low-$q$ 
maximum in the SAXS profile (Fig.~\ref{fig:profile}), manifested as a peak in 
$S\nbr*{q}$ at $q \approx 0.3~\inm$ (Fig.~\ref{fig:sq}a). 
The PCF has a primary maximum at $r = 2.2$~nm (Fig.~\ref{fig:gr}a), slightly 
less than the effective protein diameter (Table~\ref{tab:gel}). 
For comparison, in the (monomeric) crystal structure 1bpi,\cite{Parkin1996} 
there are 10 BPTI neighbours with centre-of-mass (COM) separations in the range 
2 -- 3~nm (two each at 2.26, 2.42, 2.56, 2.63 and 2.86~nm). 
Under certain conditions (high pH, high salt concentration), BPTI forms a tight 
decamer both in the crystal\cite{Hamiaux1999} and in solution.\cite{Hamiaux2000,
Gottschalk2003} 
But the pronounced minima at $q = 1.5$ and $2.9~\inm$ in the decamer form factor 
(Fig.~\ref{fig:bpti}) are not evident in our gel or solution SAXS profiles. 
We therefore conclude that decamers are not present under our conditions. 
While two BPTI molecules that are directly joined by a cross-link are expected 
to have a separation exceeding the shortest separation in the (monomeric) 
crystal, the 2.2~nm separation indicated by the PCF might be due to a tight 
approach (via the extended neutral and largely hydrophobic face of the BPTI 
molecule) of two BPTI molecules that are both cross-linked to a third one.
  
For simple liquids, the first-shell coordination number $N\rsub{c}$ is usually 
obtained by integrating the PCF up to its first minimum. 
But, for the BPTI gel, the first minimum in $g\nbr*{r}$ is very shallow and 
extended (Fig.~\ref{fig:gr}a), so we define $N\rsub{c}$ as the integral up to 
the distance $r\rsub{c} = 8.06$~nm where $g\nbr*{r} = 1$ (on the large-$r$ flank 
of the primary peak). 
This integration yields $N\rsub{c} = 31.5$ (Fig.~\ref{fig:rcn}a). 
In contrast, for a uniform protein distribution and at the same BPTI 
concentration as in the gel, we would have $N\rsub{c0} = 21.8$ at $r\rsub{c} = 8.06$~nm.
Within a sphere of radius 8.06~nm around a given BPTI molecule, the local 
protein density (concentration) defined as 
$n\rsub{P}\nbr*{r} = \sbr*{N\nbr*{r}+1}/V\nbr*{r}$, where $V\nbr*{r}$ is the 
volume of a sphere of radius $r$ centred on the reference protein molecule, is 
therefore higher than the bulk density by a factor of 
$(31.5 + 1)/(21.8 + 1) = 1.43$. 

From the shoulder seen at $r \approx 5$~nm (Fig.~\ref{fig:gr}a), it is clear 
that the primary peak in the PCF comprises 2, or even 3, overlapping 
coordination shells.
If only the first shell were included, the local density would be even higher. 
But the local density is not uniform; presumably it is higher along the 
cross-linked chains than in the regions in between. 
Indeed, in $\sim 90$~\% of the volume of the 8~nm sphere (beyond $\sim 3.5$~nm), 
the running coordination number $N\nbr*{r}$ grows more slowly than for a uniform 
distribution (Fig.~\ref{fig:rcn}a). 

In a log--log plot as in Fig.~\ref{fig:rcn}, the slope yields the fractal 
dimension, $d\rsub{f}$, defined via the scaling relation 
$N\nbr*{r} \propto r^{\,d\rsub{f}}$.\cite{Meakin1988}
Beyond $\sim 15$~nm, $N\nbr*{r}$ exhibits bulk-like scaling with $d\rsub{f} = 3$ 
(Fig.~\ref{fig:rcn}a). 
To make this more precise, we define a correlation length, $\xi$, as the 
distance where $d\rsub{f}$ has reached a value of 2.9 on its approach to the 
bulk value 3. 
Analysis of the slope in Fig.~\ref{fig:rcn}a then yields $\xi = 16$~nm for the 
BPTI gel.

%
%
\subsubsection{\label{subsubsec:mb}Mb.~~}

The SAXS profile and $S\nbr*{q}$ for the Mb gel do not show any peak at 
$q < 1~\inm$ in the examined $q$ range (Figs.\ \ref{fig:profile}b 
and~\ref{fig:sq}b). 
The PCF clearly reveals at least 3 coordination shells (Fig.~\ref{fig:gr}b) and 
remains well above ($\sim 2$) the bulk value 1 even at $r = 30$~nm 
(Fig.~\ref{fig:gr}b). 
The spatial correlations are thus of longer range in the Mb gel than in the BPTI 
gel. 
Indeed, the correlation length, $\xi = 157$~nm, is an order of magnitude longer 
than for BPTI. 

From $N\nbr*{r}$ in Fig.~\ref{fig:rcn}b, we find that the first coordination 
shell ($r < 5.0$~nm) contains 5.0, the second shell ($5.0 < r < 7.5$~nm) 9.0 and 
the third shell ($7.5 < r < 10.0$~nm) 14.5 Mb molecules. 
The first three shells ($r < 10$~nm) thus contain 28.5 Mb molecules, whereas a 
uniform distribution would only have 4.0 neighbours within 10~nm. 
This corresponds to a local density increase by a factor of $29.5/5.0 = 5.9$. 
The spatial heterogeneity is thus more pronounced in the Mb gel than in the more 
concentrated BPTI gel.
Since $S(0)$ is proportional to the mean-square fluctuation (or spatial 
variation) of the protein concentration, the stronger spatial heterogeneity in 
the Mb gel can explain the orders-of-magnitude larger $S(q)$ at 
$q \approx 0$ (Fig.~\ref{fig:sq}). 

The Mb gel analysis may not be quantitatively accurate since, for Mb, the $S(q)$ 
back-calculated from the corrected (non-negative) PCF differs somewhat from the 
original $S(q)$ (Fig.~\ref{fig:sq_back}b).
The PCF becomes negative at small $r$ (Fig.~\ref{fig:gr_orig}b) because the 
negative $S\nbr*{q}-1$ in the range $0.5 < q < 2.0~\inm$ is not fully 
compensated by a slightly positive $S\nbr*{q}-1$ at higher $q$ 
(Fig.~\ref{fig:sq}b). 
The latter feature is not resolved in the noisy high-$q$ part of the SAXS 
profile from the dilute Mb gel.

%
%
\subsubsection{\label{subsubsec:ifabp}IFABP.~~}

The SAXS profile and $S\nbr*{q}$ for the IFABP gel are qualitatively the same as 
for the Mb gel (Figs.\ \ref{fig:profile}c and~\ref{fig:sq}c).
The PCF reveals multiple coordination shells (Fig.~\ref{fig:gr}c), as for Mb, 
and it approaches the bulk value with a correlation length, $\xi = 52$~nm, 
intermediate between those for BPTI and Mb (Table~\ref{tab:gel}). 
Unlike the Mb case, the inverse transform of the non-negative $g\nbr*{r}$ 
yields a back-calculated $S\nbr*{q}$ in good agreement with the original 
$S\nbr*{q}$ (Fig.~\ref{fig:sq_back}c). 

From $N\nbr*{r}$ in  Fig.~\ref{fig:rcn}c, we find that the first coordination 
shell ($r < 4.5$~nm) contains 4.0, the second shell ($4.5 < r < 7.0$~nm) 6.9 and 
the third shell ($7.0 < r < 9.5$~nm) 7.8 IFABP molecules. 
The first three shells ($r < 9.5$~nm) thus contain 18.7 IFABP molecules, whereas 
a uniform distribution would only have 7.0 neighbours within 9.5~nm. 
This corresponds to a local density increase by a factor of $19.7/8.0 = 2.5$,
intermediate between the corresponding values for BPTI and Mb. 
As for Mb, the strong spatial heterogeneity in the IFABP gel should give rise
to a large $S(q)$ at $q \approx 0$ (Fig.~\ref{fig:sq}c).

%
%
\subsubsection{\label{subsubsec:clusters}Cluster characteristics.~~}

For all three protein gels, the position of the primary PCF peak is close to
the effective protein diameter, $\sigma\rsub{P}$ (Fig.~\ref{fig:gr}, 
Table~\ref{tab:gel}), indicating compact clusters with nearest neighbours almost 
in contact.
Because the Mb gel has a 10-fold lower protein concentration than the BPTI gel, 
the spatial heterogeneity is stronger, as indicated by the large $S(q)$ at 
$q \approx 0$.
The correlation length, $\xi$, may be regarded as a measure of the average 
cluster--cluster separation. 
The 10-fold larger $\xi$ for the Mb gel as compared to BPTI can be explained 
partly by the 10-fold lower protein concentration and partly by the larger 
clusters (Fig.~\ref{fig:gr}). 
Presumably, the Mb clusters are larger because of the larger number 
(19 versus 4) and more even distribution of lysine side-chains 
(Fig.~\ref{fig:struct}).

%
%
\subsection{\label{subsec:kinetics}Cross-linking kinetics}

\begin{figure}[!t]
  \centering
  \includegraphics[viewport=0 0 209 428]{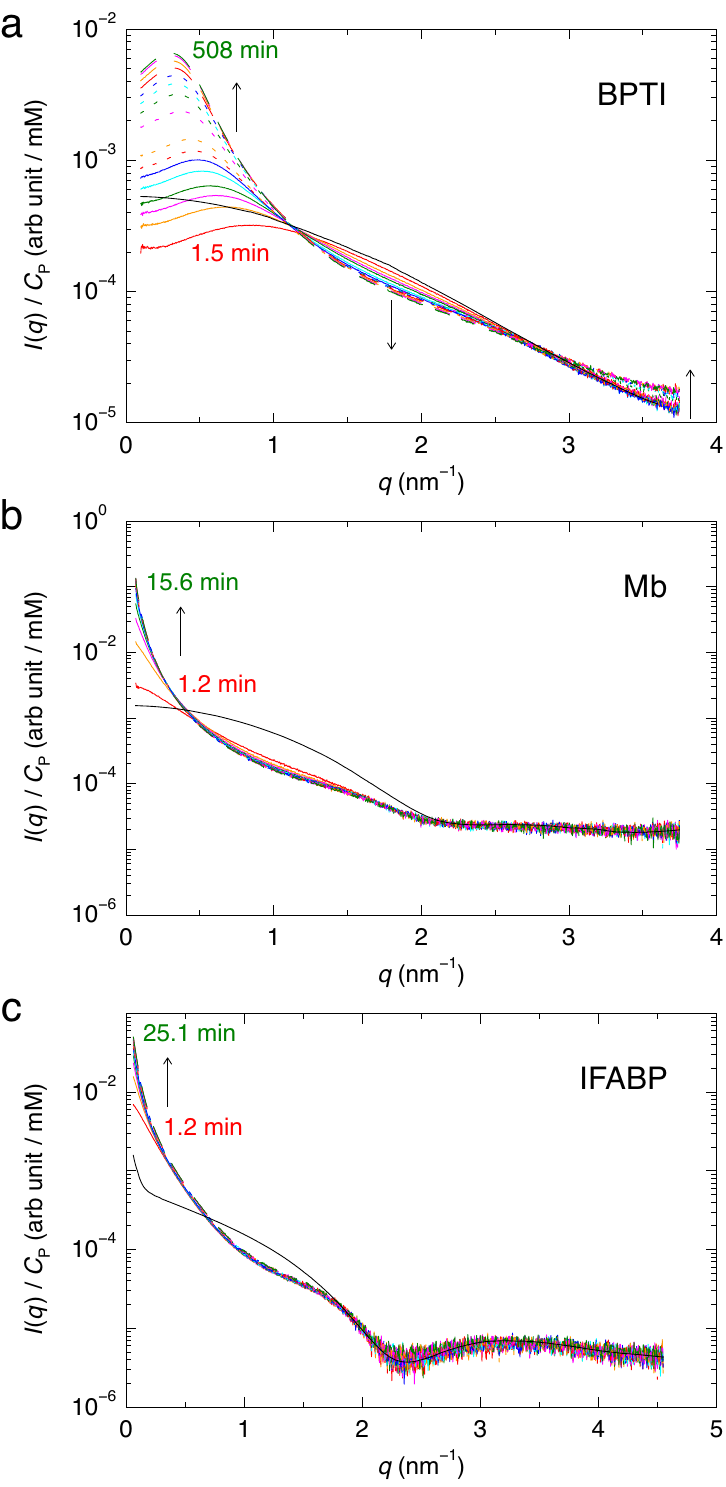}
  \caption{\label{fig:timeseries}Time-resolved SAXS profiles at 6~$\degc$ from 
           developing gels of BPTI (\textbf{a}), Mb (\textbf{b}) and IFABP 
           (\textbf{c}). 
           Time ordering is indicated by line colour (red $\rarrow$ orange 
           $\rarrow$ magenta $\rarrow$ green $\rarrow$ cyan $\rarrow$ blue), 
           line type (solid $\rarrow$ dotted $\rarrow$ dashed) and arrows. 
           The timing of the first and last profiles is given.
           The AFF is also shown for each protein (black solid curve).}
\end{figure}

To study the kinetics of gel formation by GA cross-linking, we performed 
time-resolved SAXS measurements where scattering profiles were recorded at 
regular time intervals after mixing protein and GA solutions. 
Figure~\ref{fig:timeseries} shows, for each protein, 16 profiles from the 
developing gel, along with the respective apparent form factor 
(Fig.~\ref{fig:formfactor}). 
The timing of each scattering profile, counted from the mixing of protein and GA 
solutions to the middle of the irradiation period and including a 1~min 
dead-time between mixing and the first irradiation period, is listed in 
Table~\ref{tab:kinetics}. 
Consistent with the finding that GA cross-linking alters the structure factor 
with little or no effect on the form factor (Sec.~\ref{subsec:structure}), the 
time-dependence in the SAXS profile is most evident in the low-$q$ region. 
For Mb and IFABP, due to the fast kinetics and the limited time resolution 
of the experiment, we could reliably monitor the process only at $q < 0.5~\inm$. 

From the time-resolved SAXS data, we determine the gel formation or 
cross-linking rate, $k\rsub{CL}$, by assuming first-order kinetics: 
$I\nbr*{q,\,t} = \alpha\nbr*{q} - \beta\nbr*{q}\exp\sbr*{-k\rsub{CL}(q)\,t}$. 
The $k\rsub{CL}$ values obtained from exponential fits to the time-dependent 
scattering intensity at $q = 0.1~\inm$ are shown in Table~\ref{tab:gel}.
These rates correspond to characteristic cross-linking times, 
$\tau\rsub{CL} = 1/k\rsub{CL}$, of 8.7~h, 3.3~min and 6.2~min for BPTI, Mb and 
IFABP, respectively. 

Radiation damage during the multiple irradiations enhances scattering at low 
$q$, causing the last profile in the time series to overshoot the equilibrium 
profile in Fig.~\ref{fig:profile}a.
But radiation damage only makes a minor contribution to the, mainly 
structure-related, build-up of the low-$q$ intensity and should therefore only 
give rise to a modest overestimate of $k\rsub{CL}$.  

\begin{figure}[!t]
  \centering
  \includegraphics[viewport=0 0 213 286]{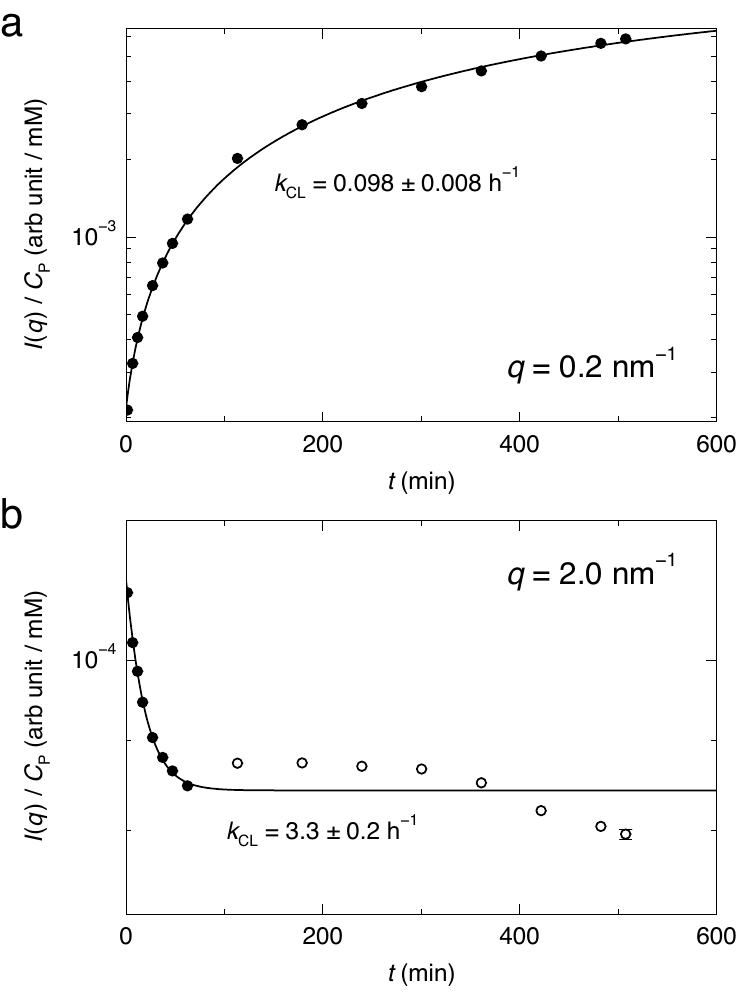}
  \caption{\label{fig:kinetics_fit} Time-dependent SAXS
  	   intensity from BPTI gel (Fig.~\ref{fig:timeseries}a) at $q = 0.2~\inm$ 
           (\textbf{a}) and $2.0~\inm$ (\textbf{b}). 
           The filled data points were used for the exponential fit.}
\end{figure}

For BPTI, $k\rsub{CL}(q)$ was determined as a function of $q$ up to $2.3~\inm$, 
except near $1~\inm$ where the time-dependent SAXS profiles show a 
quasi-isosbestic point (Fig.~\ref{fig:timeseries}a). 
The obtained $k\rsub{CL}$ values (Fig.~\ref{fig:kinetics_fit}) indicate, not 
surprisingly, that the gel structure forms faster on shorter length scales 
($2\pi/2.0 \approx 3$~nm) than on longer length scales 
($2\pi/0.2 \approx 30$~nm).
In the former case, the data suggest, in addition to the principal fast 
component, one or more minor slow components, but it cannot be excluded that 
radiation damage plays a role here.  

While we are not aware of any previous quantitative kinetic study of protein gel 
formation by GA cross-linking, the rate of formation of small cross-linked 
protein clusters has been investigated by light scattering, UV absorbance or 
ezymatic activity.\cite{Hopwood1970,Jansen1971,Tomimatsu1971,Mohapatra1994}   
At the much lower protein concentrations examined by these authors, the 
cross-linking process exhibits two distinct steps, attributed to fast cluster 
formation by cross-linking of the most reactive lysines followed by slower 
linkage of clusters.\cite{Tomimatsu1971} 
At the higher protein concentrations studied here, the time scales of these two 
processes may overlap, leading to an apparently exponential build-up of 
scattering intensity at low $q$ (Fig.~\ref{fig:kinetics_fit}a).

The rate of gel formation depends on many factors, including protein and GA 
concentrations, pH, temperature and availability of primary amino groups. 
While it is outside the scope of this study to systematically explore these 
factors, we note that the BPTI gel forms 2 orders of magnitude slower than the 
Mb and IFABP gels.
The BPTI gel differs from the two other gels in having a much higher protein 
concentration (Table~\ref{tab:gel}). 
But this should accelerate gel formation, so there must be other factors at 
play. 
We suspect that the dominant factor here is the $\sim$ 3 units lower pH in the 
BPTI gel (Table~\ref{tab:gel}), which means that the fraction of 
\textepsilon-amino groups in the reactive deprotonated (\ce{NH2}) 
form\cite{Walt1994,Migneault2004} is 3 orders of magnitude lower in the BPTI 
gel.
Other, presumably less important, effects of a low pH include suppression of GA 
aldol condensation,\cite{Walt1994,Migneault2004,Hardy1969,Korn1972,Whipple1974,
Margel1980,Tashima1991,Kawahara1992} which might lead to shorter cross-links and 
slower gel formation, and a more positive net protein charge, $Z$ 
(Table~\ref{tab:gel}).
A larger $\abs*{Z}$ should retard gel formation and make the clusters smaller 
due to intra-cluster Coulomb repulsion, but, since cross-links remove positive 
lysine charges, it is not clear which of the three proteins has the largest 
$\abs*{Z}$.
Yet another factor that could slow down BPTI gel formation is the small number 
(Table~\ref{tab:gel}) and uneven distribution (Fig.~\ref{fig:struct}) of lysine 
side-chains in BPTI, which may also be responsible for the smaller size of the 
BPTI clusters (Sec.~\ref{subsubsec:clusters}).    

%
%
\section{\label{sec:conc}Conclusions}

The SAXS data presented here provide quantitative information about 
microstructure and aggregation kinetics in GA cross-linked gels of three 
different proteins. 
While the three protein gels are qualitatively similar, the BPTI gel differs 
quantitatively from the Mb and IFABP gels. 
This difference is caused by a combination of factors, among which the most 
important are the higher protein concentration, the lower pH and the more 
limited availability of lysine amino groups in the case of BPTI. 
The most important conclusions derived from our analysis of the SAXS data are as 
follows. 

\textit{The native protein structure is retained in the cross-linked gel}, at 
least at the $\sim 1$~nm resolution afforded by the SAXS data. 
This conclusion follows from the close agreement at high $q$ between the SAXS 
profiles from gel and dilute solution (Fig.~\ref{fig:profile}).
The evidence is most clear-cut for Mb and IFABP, whereas, for the smaller 
protein BPTI, the form factor is masked by $S(q)$ oscillations that persist to 
higher $q$. 
While the lysine side-chains involved in cross-links, and perhaps some nearby 
side-chains as well, must be conformationally perturbed, the SAXS data rule out 
a significant degree of unfolding. 
This conclusion is consistent with the finding, from MRD measurements on 
BPTI,\cite{Persson2008a} that the ns -- \us\ dynamics of internal water 
exchange, and the rate-limiting structural fluctuations,\cite{Persson2013} are 
essentially the same whether the protein is free in solution or cross-linked in 
a gel. 
Furthermore, the SAXS results indicate that this is true also for Mb and IFABP.        

\textit{The protein gel is spatially heterogeneous, with dense clusters 
linked by sparse networks.} 
The strong spatial heterogeneity in the more dilute Mb and IFABP gels produces 
intense scattering at low $q$. 
The BPTI gel, with a higher concentration of smaller clusters and a shorter 
correlation length, is less strongly heterogeneous. 
The low-$q$ scattering is therefore much weaker, allowing us to observe a peak 
at $q \approx 0.3~\inm$, resulting from the interplay of intra-cluster 
attraction and inter-cluster repulsion.

\textit{Within the clusters, adjacent protein molecules are almost in contact}. 
The number of nearest protein neighbours, estimated by integrating over the 
first coordination shell (not perfectly well-defined for BPTI) in the PCF 
(Fig.~\ref{fig:gr}), is $5 \pm 1$ for all three proteins. Since this exceeds the 
value of 2 expected for a linear chain, the protein clusters must be multiply 
connected. 
Some nearest neighbours may be cross-linked via a third protein molecule while 
still approaching each other almost to contact via hydrophobic attraction. 
Despite such close encounters, the cluster is not uniformly dense. 
For the BPTI gel, the protein concentration within 8~nm of a reference molecule 
(a spherical volume that includes 2 or 3 coordination shells) is $\sim 24$~\mM\ 
(43~\% above the average concentration in the gel), which is still much lower 
than the concentration of $136$~\mM\ in the (monomeric) BPTI crystal 
1bpi.\cite{Parkin1996} 
Yet, the close protein encounters that do occur within a cluster should lead to 
a stronger dynamical perturbation of water dynamics than in the hydration layer 
of a protein in dilute solution. 
At points of particularly close contact between two protein molecules, some 
water molecules may be trapped with survival times exceeding 1~ns, as seen for 
internal water molecules. 
Both of these phenomena, enhanced dynamical perturbation in the hydration layer 
and trapped water molecules with survival times in the range 1 -- 10~ns, have 
been inferred from MRD studies of cross-linked proteins.\cite{Persson2008a,
Kaieda2013a}
   
\textit{Proteins with a large number of uniformly distributed lysine side-chains 
make larger clusters}.
This generalisation is based on the correlation between the range of 
protein--protein correlations, as reflected in the PCF (Fig.~\ref{fig:gr}), with 
the number (Table~\ref{tab:gel}) and surface distribution 
(Fig.~\ref{fig:struct}) of lysine amino groups in the three investigated 
proteins. 
The correlation length, $\xi$, defined in terms of the fractal dimension, is 
more closely related to the typical cluster--cluster separation and therefore 
depends on the overall protein concentration as well as on the cluster size.

\textit{Gel formation occurs on time scales from minutes to hours} under our 
conditions (Fig.~\ref{fig:timeseries}).
As judged by the scattering intensity at $q \le 0.2~\inm$, gel formation appears 
to obey first-order kinetics (Fig.~\ref{fig:kinetics_fit}). 
The much slower gel formation for BPTI is attributed to the lower pH and the 
consequent lower abundance of reactive deprotonated amino groups.

%
%
\section*{Acknowledgements}

{\small
We thank Hanna Nilsson for Mb purification, Annika Rogstam at Lund Protein 
Production Platform for IFABP preparation, Bayer Healthcare AG for a generous 
supply of BPTI and MAX-lab Synchrotron beamline I911-4 for beamtime under 
proposal ID 20120020.
This work was financially supported by a project grant (to B.H.) from the 
Swedish Research Council and postdoctoral fellowships (to S.K.) from the 
Wenner-Gren Foundations and the Swedish Research Council.
}

%
%

%
%
\onecolumn
\section*{Supporting information}

\setcounter{section}{0}
\setcounter{figure}{0}
\setcounter{table}{0}
\setcounter{equation}{0}

\renewcommand{\theHsection}{sup.\the\value{section}}
\renewcommand{\thesection}{S\arabic{section}}
\renewcommand{\thefigure}{S\arabic{figure}}
\renewcommand{\thetable}{S\arabic{table}}
\renewcommand{\theequation}{S\arabic{equation}}

\vspace*{\fill}
\begin{figure}[!h]
  \centering
  \includegraphics[viewport=193 207 402 635]{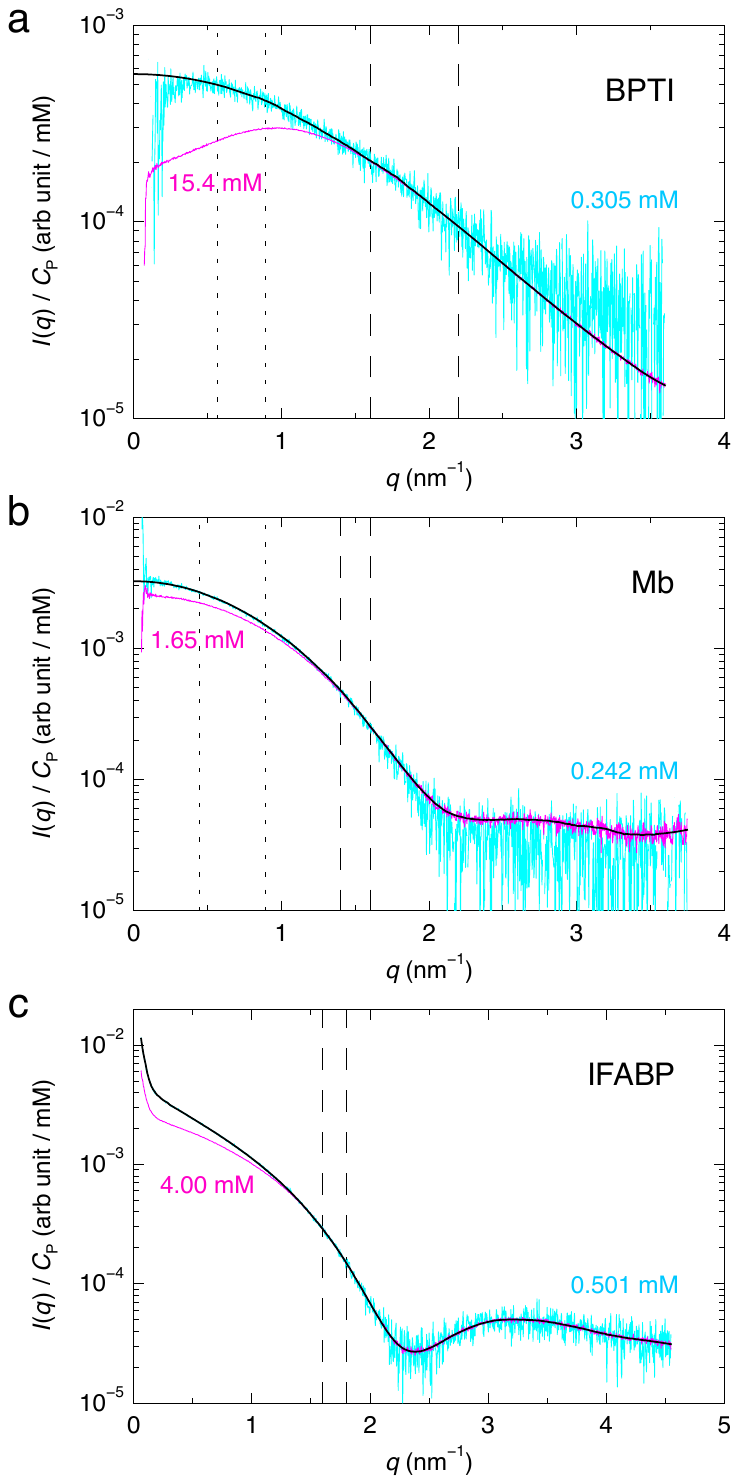}
  \caption{\label{fig:formfactor}Apparent form factor (black) and solution SAXS 
           profiles (magenta and cyan) for BPTI (\textbf{a}), Mb (\textbf{b}) and 
           IFABP (\textbf{c}).
           Dashed and dotted vertical lines indicate the $q$ windows where the two 
           solution scattering files were superimposed and where the Guinier 
           analysis was performed, respectively (Sec.~\ref{subsec:saxsanal}).}
\end{figure}
\vspace*{\fill}

\newpage

\vspace*{\fill}
\begin{figure}[!h]
  \centering
  \includegraphics[viewport=194 207 401 635]{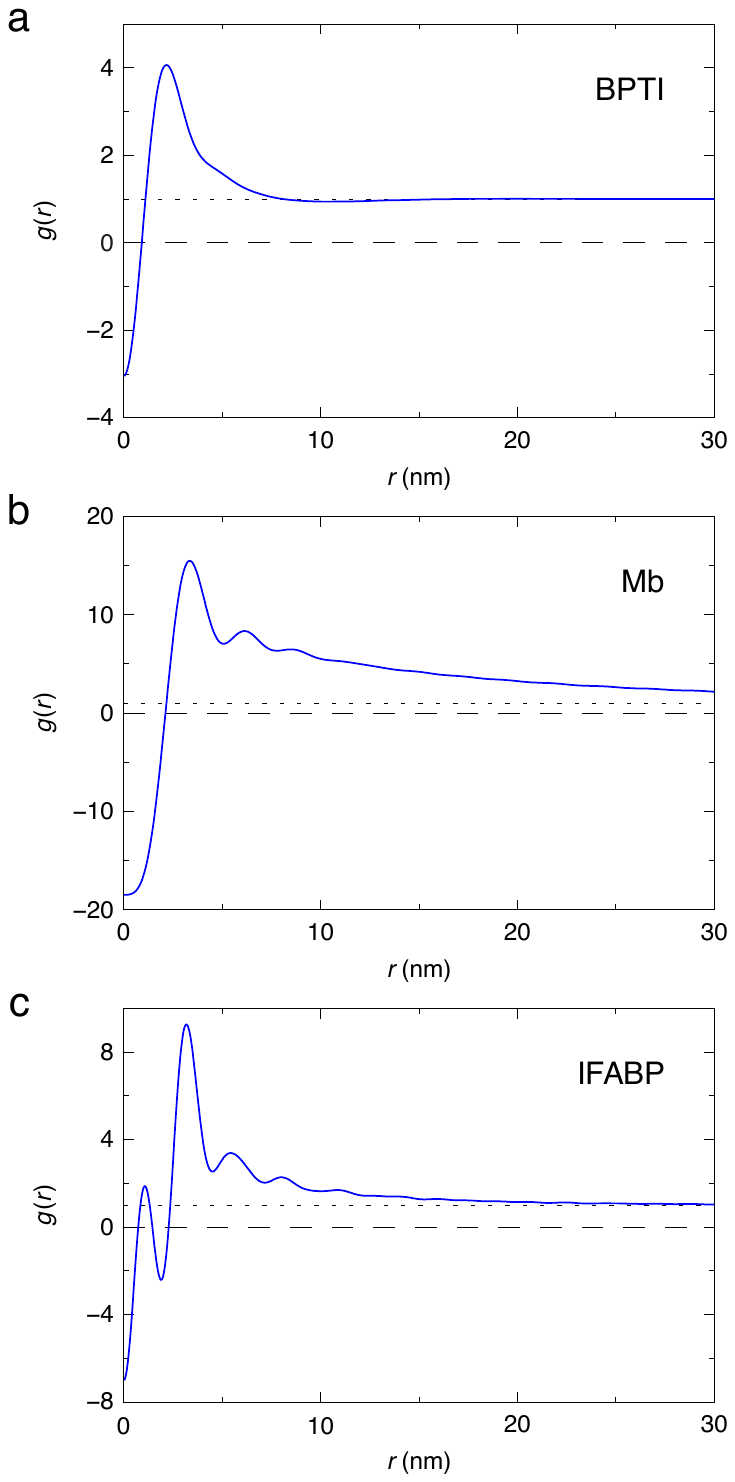}
  \caption{\label{fig:gr_orig}Pair correlation function, $g\nbr*{r}$, for 
           cross-linked BPTI (\textbf{a}), Mb (\textbf{b}), and IFABP 
           (\textbf{c}), obtained by fast sine transform of the modified 
           structure factor, $S\nbr*{q}$ (blue curve in Fig.~\ref{fig:sq}). 
           The dotted line indicates the asymptote $g(r\rightarrow\infty) = 1$. 
           The negative $g\nbr*{r}$ at small $r$ is an artifact caused by setting
           $S(q) = 1$ at high $q$.}
\end{figure}
\vspace*{\fill}

\newpage

\vspace*{\fill}
\begin{figure}[!h]
  \centering
  \includegraphics[viewport=194 207 401 635]{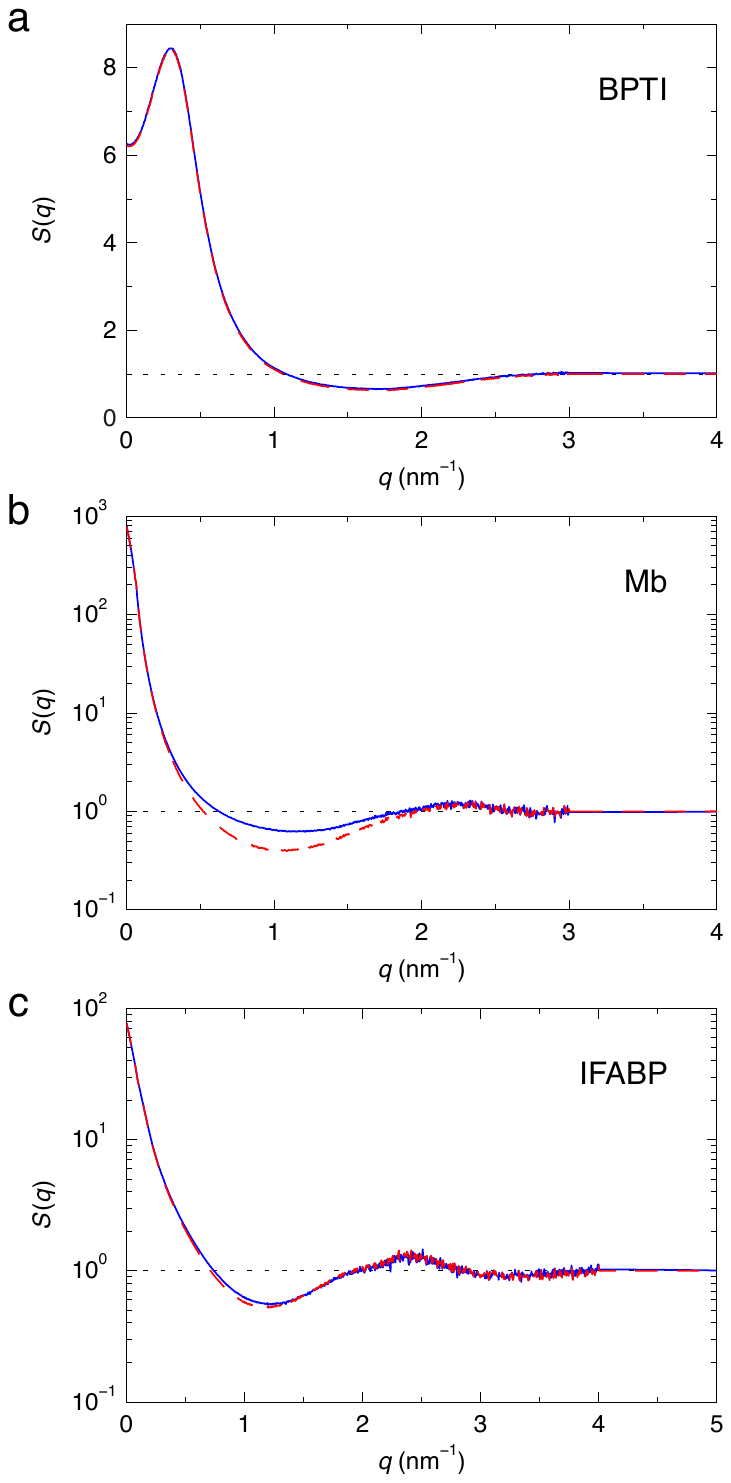}
  \caption{\label{fig:sq_back}Back-calculated structure factor, $S\nbr*{q}$, for 
           cross-linked BPTI (\textbf{a}), Mb (\textbf{b}) and IFABP (\textbf{c}). 
           The red dashed curve is the inverse sine transform of the modified 
           (non-negative) $g\nbr*{r}$. 
           The blue curve is the original $S\nbr*{q}$ (blue curves in 
           Fig.~\ref{fig:sq}).}
\end{figure}
\vspace*{\fill}

\newpage

\vspace*{\fill}
\begin{figure}[!h]
  \centering
  \includegraphics[viewport=194 349 401 493]{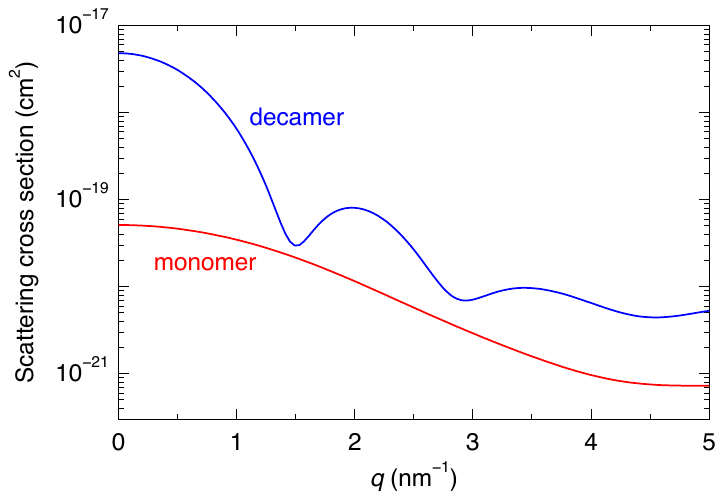}
  \caption{\label{fig:bpti}Scattering cross section for  BPTI monomer (red) and 
           decamer (blue). 
           The curves were calculated from the atomic coordinates of the crystal 
           structures 1bpi\cite{Parkin1996} (monomer) and 1bhc\cite{Hamiaux1999} 
           (decamer) using the program \textsc{crysol}\cite{Svergun1995} without 
           hydration-layer correction.}
\end{figure}

\vspace*{\fill}

\begin{table}[!h]
  \centering
  \small
  \begin{threeparttable}
    \caption{\label{tab:kinetics}Timing of kinetics series.$\rsup{a}$}
    \begin{tabular*}{7.5cm}{@{\extracolsep{\fill} }lrrr}
      \toprule
      & \multicolumn{3}{c}{$t$ (min)} \\
      \cmidrule{2-4}
      \# & BPTI\tnote{\,b} & Mb\tnote{\,c} & IFABP\tnote{\,c} \\
      \midrule
      1  &   1.5 &  1.2 &  1.2 \\[2pt]
      2  &   6.8 &  2.2 &  2.8 \\[2pt]
      3  &  11.8 &  3.1 &  4.3 \\[2pt]
      4  &  16.9 &  4.1 &  5.9 \\[2pt]
      5  &  27.0 &  5.1 &  7.4 \\[2pt]
      6  &  37.2 &  6.0 &  8.9 \\[2pt]
      7  &  47.3 &  7.0 & 10.5 \\[2pt]
      8  &  62.6 &  7.9 & 12.0 \\[2pt]
      9  & 113.2 &  8.9 & 13.6 \\[2pt]
      10 & 178.9 &  9.8 & 15.1 \\[2pt]
      11 & 239.7 & 10.8 & 16.6 \\[2pt]
      12 & 300.3 & 11.7 & 18.2 \\[2pt]
      13 & 361.1 & 12.7 & 19.7 \\[2pt]
      14 & 421.7 & 13.6 & 22.0 \\[2pt]
      15 & 482.5 & 14.6 & 23.6 \\[2pt]
      16 & 507.9 & 15.6 & 25.1 \\ 
      \bottomrule
    \end{tabular*}
    \footnotesize
    \begin{tablenotes}[flushleft]
      \item[a] Time elapsed from GA addition to the middle of the irradiation 
               period, including a 1~min dead time.
      \item[b] 60~s irradiation period.
      \item[c] 10~s irradiation period.
    \end{tablenotes}
  \end{threeparttable}
  \normalsize
\end{table}
\vspace*{\fill}

\end{document}